# Breaking optoelectronic SNR limitations via physics-consistent computational diffractive imaging

Yun Xie[1], Bianli Zhao[2], Han Yue[1], Rui Zhang[2], Zhiyi Huang[3], Weiran Jiang[3], Chuangchuang Cheng[4], Steve F. Shu[3,*]

1. College of Intelligent Robotics and Advanced Manufacturing, Fudan University, Shanghai, 200433, P.R. China

2. Department of Materials and Energy, Yunnan University, Kunming 650504, P.R. China

3. School of Electrical and Computer Engineering, The University of Sydney, Camperdown, NSW 2006, Australia

4. Anhui Province Key Laboratory of Measuring Theory and Precision Instrument, Hefei University of Technology, Hefei, Anhui 230009, P.R. China

* Address correspondence to: steve.shu@sydney.edu.au

**Abstract**

Ptychography is a powerful lensless imaging technique capable of approaching the diffraction limit, yet its performance is increasingly constrained by non-ideal detection hardware. In photon-limited measurements, weak high-frequency diffraction signals often overlap with spatially heterogeneous detector noise, whereas most reconstruction algorithms still treat the detector as an ideal measurement plane. Here, we introduce detector-informed measurement consistency into ptychographic reconstruction. By calibrating the pixelwise sensor response, the method construct a spatially resolved confidence map and embed it into the iterative amplitude constraint, allowing unreliable detector residuals to be down-weighted while preserving physically meaningful diffraction information. Experiments across transmission, reflection, and weak biological phase imaging show improved diffraction-data quality, an approximately twofold signal-to-noise ratio (SNR) enhancement, and reconstruction approaching the Rayleigh limit with a measured (k)-factor of about 0.65. Compared with previous advanced denoising methods, the proposed framework achieves a better balance between suppressing detector-induced background and preserving structural diffraction information. These results show that detector reliability can be used as an in-loop physical constraint to extend the performance of ptychographic imaging with imperfect sensors.

## 1 Introduction

The quantitative recovery of complex optical fields is essential for understanding light–matter interactions (Park et al. 2018; Wang et al. 2022; Nguyen et al. 2022; Park et al. 2023), especially when phase information is needed to reveal weak scattering, refractive-index variations, or fine structural contrast (Chapman and Nugent 2010; Ozcan and McLeod 2016; Pfeiffer 2018a; Valzania et al. 2019). Conventional optoelectronic detectors, however, record only intensity and therefore discard the phase of the optical wavefront. Ptychography addresses this limitation by reconstructing the complex field from a set of redundant, overlapping diffraction measurements (Jagatap and Hegde 2019; Bangun et al. 2022; Huang et al. 2015; Huijts et al. 2020; Zheng et al. 2013; Tanksalvala et al. 2021). Since the reconstruction relies on the numerical inversion of measured diffraction intensities, ptychography is particularly sensitive to data fidelity and provides a useful model system for examining how physical measurements enter computational reconstruction. This has enabled its broad use in visible-light (Maiden et al. 2012; A. Maiden et al. 2017), extreme-ultraviolet (EUV) (Yang et al. 2022; Karl et al. 2018), X-ray (Clark et al. 2012; Pfeiffer 2018b; Thibault et al. 2008), and electron (Dong et al. 2025, 2024) imaging, with applications ranging from semiconductor metrology (Gardner et al. 2017) to biological characterization (Loetgering et al. 2020; Maiden et al. 2012; A. Maiden et al. 2017; Odstrcil et al. 2016).

Despite these advances, the fidelity and effective resolution of ptychographic imaging remain strongly affected by the non-ideal response of detection hardware. In many frontier imaging scenarios, the available photon flux is limited by source efficiency, exposure time, or sample-damage thresholds, making weak high-frequency diffraction signals especially vulnerable to noise. Practical detectors also introduce photon shot noise, readout fluctuations, thermal noise, dark-current drift, and pixel-dependent response variations (Clark et al. 2012; Liu et al. 2024; Tan et al. 2022; Liu et al. 2023; Thibault and Guizar-Sicairos 2012; Cheremkhin et al. 2021; Ye et al. 2021). Most reconstruction engines simplify this measurement process by treating the detector as an ideal plane on which every recorded intensity value provides an equally reliable constraint. This assumption separates detector physics from the inverse problem. As a result, the iterative update cannot readily distinguish meaningful diffraction information from unreliable detector responses. In low-flux regimes such as EUV and X-ray imaging, this mismatch can become a major factor limiting the signal-to-noise ratio (SNR) and the recoverable spatial-frequency range (Wiedorn et al. 2017; Reinhardt et al. 2017). Our previous study also showed that suppressing background fluctuations is essential for stable high-resolution reconstruction (Xie, Lin, et al. 2025). These observations suggest that the detector should not be viewed only as a passive recording plane, but also as a physical component that defines the reliability of the measured diffraction constraints.

Several strategies have been developed to account for noise and detector imperfections in phase retrieval and computational imaging. A common approach is to introduce a parametric description of the measurement noise, such as constant offsets (Thurman and Fienup 2009), Gaussian or Poisson models (Thibault and Guizar-Sicairos 2012; Godard et al. 2012), likelihood-based formulations (Thibault and Guizar-Sicairos 2012), or variance-stabilizing transforms (Zhang et al. 2017). These methods provide useful statistical regularization and can describe many experimental measurements when the corresponding noise parameters are properly calibrated. However, in practical reconstruction, simplified global parameterizations or fixed calibration conditions may not fully describe detector responses that vary across pixels and exposure settings. Explicit background terms can also bias the reconstruction when the assumed model deviates from the actual detector response. Consequently, even adaptive constraints (Konijnenberg et al. 2018) may suppress detector artifacts only at the risk of weakening faint diffraction information.

Another line of work improves the measured diffraction data before reconstruction. Local minimization, thresholding, dark-frame subtraction (DFS), and adaptive filtering (Wang et al. 2017; Qiao et al. 2023) can reduce visible background fluctuations and improve the apparent SNR. However, these methods treat the diffraction measurement as an image to be cleaned before it enters the inverse problem. Once the data have been modified outside the reconstruction loop, the physical solver can no longer distinguish original diffraction signals from artificially altered measurements. This out-of-loop processing may remove weak high-frequency components or introduce irreversible distortions. Recent in-the-loop and feature-aware strategies have begun to address this limitation by incorporating additional information into the phase-retrieval objective (Weber et al. 2024; Zhang et al. 2025; Seifert et al. 2023). The present work follows this broader direction, but addresses a different source of mismatch, the spatially nonuniform reliability of the measured diffraction constraints imposed by the detector itself.

Here, we introduce detector-informed measurement consistency for high-fidelity ptychographic reconstruction with imperfect detection hardware. Rather than treating all detector pixels as equally reliable, the proposed framework applies measured diffraction constraints according to the calibrated reliability of the detection process. A spatially resolved confidence map is constructed from sensor-response characteristics and embedded into the iterative amplitude-projection step, where it directly modulates the detector-plane residual entering the reconstruction update. In this way, unreliable detector responses contribute less to the iterative correction, while high-confidence diffraction information is retained within the physical forward model. We validate this confidence-weighted formulation in transmission and reflection ptychographic experiments involving different detector architectures, wavelengths, imaging geometries, and sample types. Compared with DFS, minimization-based background removal (MBR) as a preprocessing strategy, and in-loop Poisson–Gaussian maximum likelihood (PG-ML) reconstruction, the proposed method provides a more balanced suppression of detector-induced background and preservation of structural diffraction information. The results show improved diffraction-data quality, an approximately twofold SNR enhancement, and reconstruction approaching the Rayleigh diffraction limit, with a measured (k)-factor of approximately 0.65. The method also reduces anisotropic artifacts in reflection geometry and improves tissue-background phase contrast in weakly scattering biological samples. In addition, using the 500-iteration reconstruction as the reference, the confidence-weighted update reaches a near-converged reconstruction after approximately 300 iterations, corresponding to about 40% fewer iterations required to obtain comparable image quality. These results indicate that detector reliability can serve as an effective in-loop measurement constraint for robust ptychographic imaging with non-ideal detection hardware.

# 2 Results

## 2.1 Experimental setup

Fig. 1a shows the schematic of the experimental setup. Quasi-monochromatic illumination is provided by a semiconductor laser. The beam is attenuated with neutral density filters (ND1, ND2) and then focused onto a pinhole by a 50-mm focal length lens (L1). An imaging lens with a 40 mm focal length (L2) images the pinhole to produce a probe of 50 μm diameter. This focused probe interacts with the sample in either transmission or reflection geometry, generating a coherent diffracted wavefield. Sample movement follows the spiral scanning trajectory shown in Fig. 1b and is executed using a linear translation stage. To ensure sufficient redundancy between scan positions, a 10-μm step size is used to yield approximately 80% overlap between adjacent positions. The diffraction patterns are recorded by a detector, which is positioned at different downstream locations depending on the detection geometry.

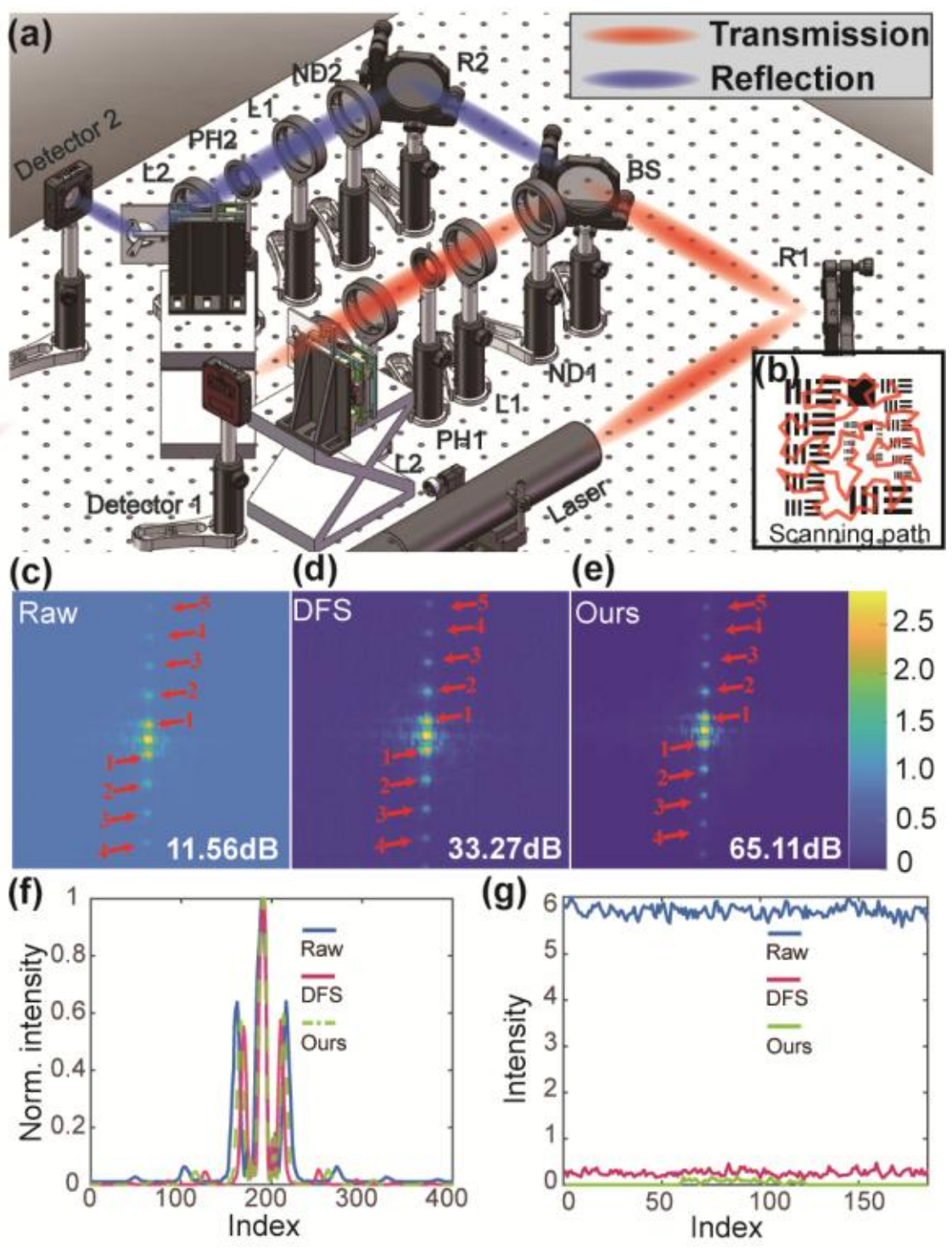

*Fig. 1 Experimental setup and analysis. (a) Schematic of the ptychographic imaging system. (b) Scanning path executed by the motion stage on the sample. (c)–(e) Diffraction patterns generated using different methods, with one representative frame shown. (f)–(g) Intensity statistics of all diffraction orders and background noise within the diffraction patterns.*

## 2.2 Resolution analysis in the near-diffraction-limited regime

To experimentally benchmark the Rayleigh resolution limit, a transmission measurement was performed using a USAF-1951 test target (HIGHRES-1, Newport) under coherent illumination with a wavelength of 455 nm. Diffraction patterns were recorded by a detector (HD-R1201M-GigE, Daheng Optics) with a pixel size of 1.85 μm and a resolution of $4000 \times 3000$ pixels, and subsequently cropped to $3000 \times 3000$ pixels for analysis. A short camera integration time of 2 ms was employed, which is considerably shorter than the exposure times used in the noise-characterization measurements. The detector was positioned 6.80 mm downstream of the sample plane. Under this configuration, the optical system achieved a numerical aperture (NA) of 0.396. To balance NA coverage against computational efficiency, the recorded diffraction patterns were further binned to $1024 \times 1024$ pixels prior to reconstruction.

Figs. 1c–e compare diffraction patterns obtained with different correction strategies. Fig. 1c shows the raw measurement without correction, where the diffraction signal is strongly embedded in background fluctuations. The reported SNR was estimated using a background-statistics metric, defined by comparing the mean squared

intensity of the full diffraction frame with that of a selected signal-free background region. The same background region was used for all compared methods. Based on this definition, the raw measurement gives an SNR of 11.56 dB. Fig. 1d presents the result after DFS, which increases the background-based SNR to 33.27 dB. This increase reflects the reduction of residual background power in $I_{\mathrm{bg}}$, rather than an increase in the intrinsic photon-limited signal. DFS is an out-of-loop correction that subtracts a background frame from $I_j^{\mathrm{mea}}(\boldsymbol{q})$ before reconstruction and then uses the corrected intensity as the input data. Although this procedure improves the overall diffraction-to-background contrast, it does not account for the spatially varying reliability of detector pixels. As a result, several weak high-order diffraction features remain affected by residual background fluctuations. This is particularly evident for the fourth and fifth diffraction orders, where the signal intensity is lower and the local diffraction-to-background contrast remains insufficient for stable reconstruction.

In contrast, the proposed method shown in Fig. 1e increases the SNR to 65.11 dB, corresponding to improvements of 53.55 dB over the raw input and 31.84 dB over DFS. This quantitative enhancement highlights the superior noise-separation capability of the proposed framework. The constructed feature domain further disentangles diffraction features from shot noise, as illustrated in Figs. 1f,g. Spectral analysis across all diffraction orders in Fig. 1f shows consistent envelopes for all three methods, confirming that diffracted intensity is faithfully preserved. More importantly, Fig. 1g demonstrates that the proposed method strongly suppresses residual noise fluctuations, confining them to values near zero, which is critical for accurate reconstruction of fine structural details.

Fig 2 summarizes the experimental resolution characterization. The comparison between DFS and the proposed framework is shown in Fig. 2a,b. DFS retains the coarse layout of the USAF target, but the reconstructed fine features are affected by residual background fluctuations and local artifacts. In contrast, the proposed method provides clearer separation of the line-pair structures in ROI 1, as shown by the magnified insets. The local intensity profiles in Fig. 2c further quantify this difference. Across the selected feature, the proposed reconstruction exhibits a steeper transition and a lower residual floor than DFS, indicating improved contrast at the resolved edge. Using the 90% intensity-drop criterion, the corresponding feature width is estimated to be approximately 910 nm. To avoid relying only on local profile measurements, Fourier ring correlation was also used for resolution assessment. As shown in Fig. 2d, the fourier ring correlation (FRC) curve crosses the 1/2-bit threshold at a spatial frequency corresponding to a resolution of 745 nm. With the experimental numerical aperture, this gives a $k$-factor of approximately 0.65, close to the Rayleigh criterion of 0.61.

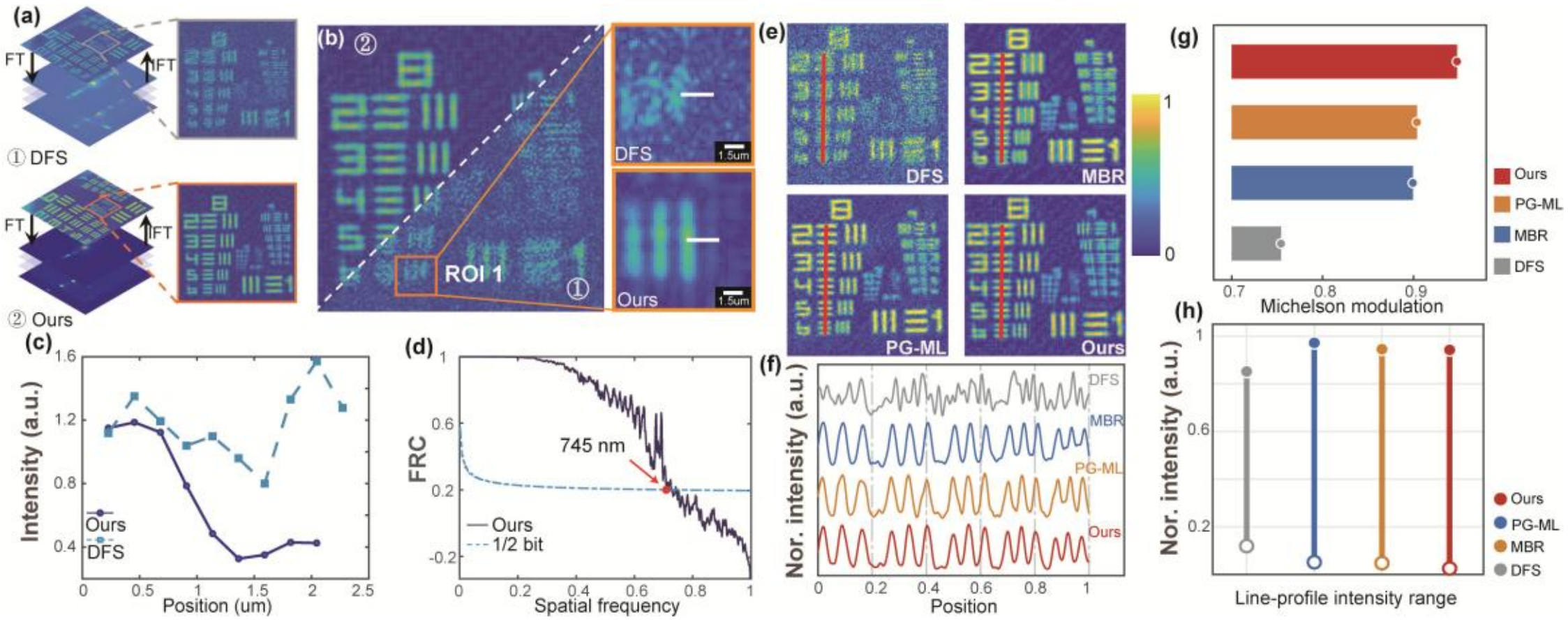

*Fig. 2 Experimental resolution characterization and comparison with representative denoising baselines. (a) Schematic comparison of reconstruction workflows using dark-frame subtraction (DFS) and the proposed detector-informed framework. (b) Reconstructed USAF target obtained with the proposed method, with ROI 1 selected for local resolution analysis. The magnified insets compare the same region reconstructed by DFS and the proposed method. (c) Line profiles extracted from ROI 1. (d) Fourier ring correlation (FRC) analysis of the proposed reconstruction. (e) Reconstructed USAF targets obtained with DFS, minimization-based background removal (MBR), Poisson–Gaussian maximum likelihood (PG-ML), and the proposed method. (f) Intensity-profile analysis extracted from the marked region in (e). (g) Michelson modulation calculated from the selected USAF line-pair profiles. (h) Peak-to-valley intensity range of the line profiles used for the modulation calculation.*

We further compared the proposed method with three representative baselines, including DFS, MBR as a preprocessing method, and PG-ML as a prescribed-noise-model in-the-loop reconstruction method. The reconstructed USAF targets are shown in Fig. 2e. DFS leaves strong high-frequency fluctuations around the target, whereas PG-ML suppresses part of the background at the cost of reduced line-pair contrast. MBR provides improved background removal, but residual nonuniformity remains visible in the reconstructed target. The proposed method maintains well-separated USAF bars while reducing the surrounding background fluctuations. This trend is also reflected in the line-profile comparison in Fig. 2f. The proposed reconstruction preserves a large peak-to-valley separation over repeated line pairs, while PG-ML shows a weaker modulation. The Michelson modulation analysis in Fig. 2g gives the highest value for the proposed method, confirming that the selected line pairs are most clearly separated in this reconstruction. The corresponding peak-to-valley ranges in Fig. 2h show that the improvement arises from both a higher retained signal level and a lower residual minimum, rather than from a simple rescaling of the intensity.

## 2.3 Mitigating anisotropic resolution in reflection geometry

To examine the applicability of the proposed framework beyond transmission geometry, we further performed reflection-mode ptychographic imaging using a 635 nm laser source. In this configuration, a semiconductor sample was illuminated at an incidence angle of 60°, and the scattered wavefronts were recorded by a high-resolution complementary metal-oxide-semiconductor (CMOS) camera (QHY600PH, QHYCCD, $9600 \times 6422$ pixels, 3.76 μm pixel size) positioned 15.70 mm below the sample, as shown in Figs. 3a,b. The recorded diffraction frames were cropped to $1024 \times 1024$ pixels for reconstruction. This experiment used exposure times beyond the calibration range of the photon-free detector characterization, thereby testing the extrapolation capability of the detector-informed weighting strategy.

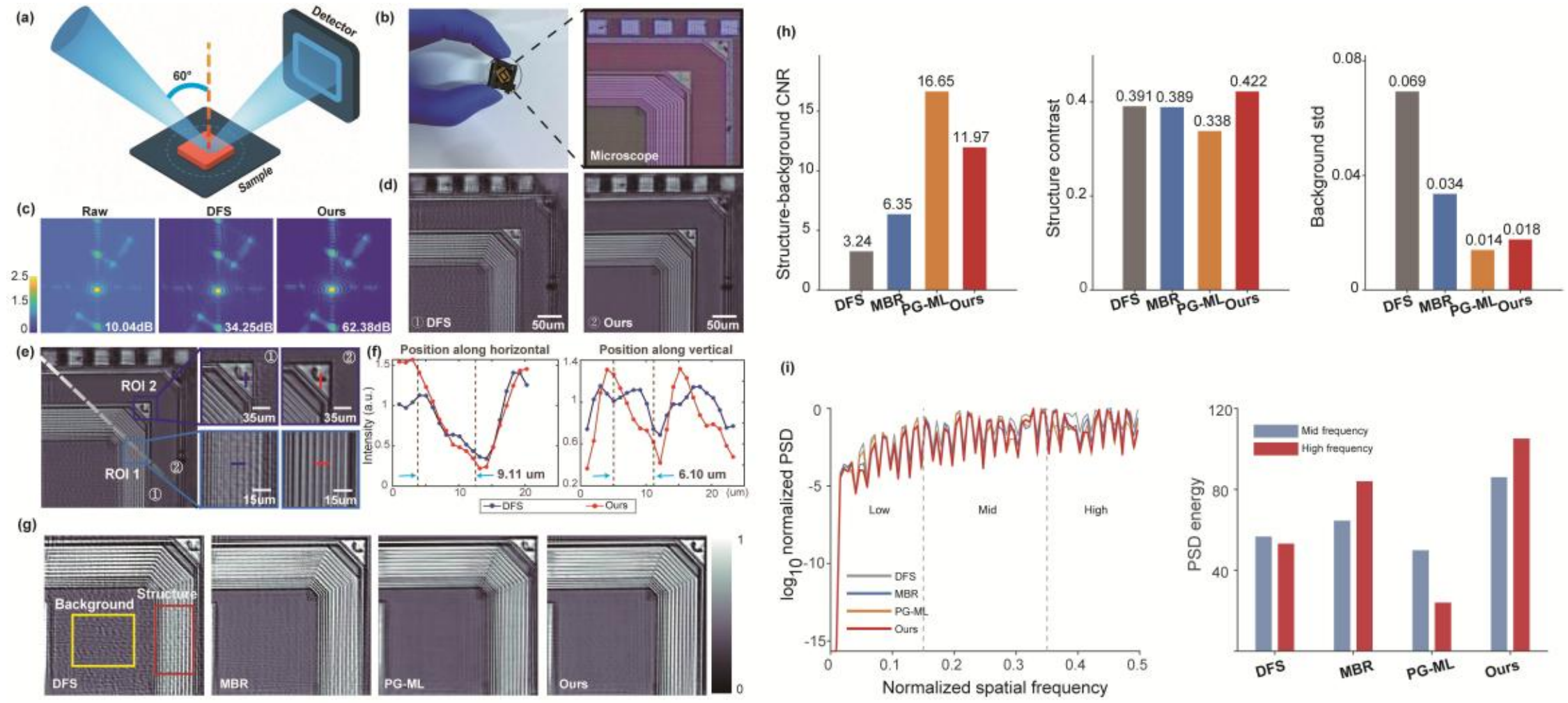

*Fig. 3 Reflection-mode experimental validation. (a) Schematic of the reflection-mode imaging geometry. (b) Semiconductor sample and microscope image of the measured structure. (c) Representative diffraction patterns obtained from the raw measurement, DFS, and the proposed method. (d) Reconstructions obtained with DFS and the proposed method. (e) Selected regions of interest used for anisotropic-resolution analysis. (f) Line-profile analysis along the horizontal and vertical directions. (g) Reconstructions obtained with DFS, MBR, PG-ML, and the proposed method. (h) Quantitative comparison of structure-background contrast-to-noise ratio (CNR), structural contrast, and background standard deviation. (i) Radially averaged power spectral density (PSD) analysis and mid-/high-frequency PSD energy comparison.*

Fig 3c compares the diffraction patterns obtained from the raw measurement, DFS, and the proposed method. The raw diffraction pattern is strongly affected by background noise, while DFS reduces the overall background level but leaves residual nonuniform fluctuations. The proposed method increases the diffraction-pattern SNR from 10.04 dB for the raw measurement and 34.25 dB for DFS to 62.38 dB. The diffraction patterns also show an asymmetric distribution of diffraction orders, which is expected in the reflection geometry because of the large incidence angle and the tilted-plane correction. The influence of this asymmetry is visible in the reconstructions shown in Fig. 3d. With DFS, the horizontal structures are largely retained, whereas the vertical etched lines are more strongly affected by blur and residual artifacts. The reconstruction obtained with the proposed method shows more uniform visibility across both directions. To quantify this behavior, two regions of interest were selected, as shown in Fig. 3e. The line profiles in Fig. 3f indicate that the proposed method provides clearer intensity transitions along both the horizontal and vertical directions, with resolvable feature widths of 9.11 μm and 6.10 μm, respectively.

We further compared the proposed method with three representative baselines. The corresponding reconstructions are shown in Fig. 3g. DFS retains substantial background texture around the patterned region. MBR reduces part of this background, but residual structural nonuniformity remains visible. PG-ML gives strong background suppression, yet the reconstructed etched features appear smoother. The proposed method preserves the line structures while reducing the surrounding background fluctuations.

The quantitative results in Fig. 3h show the different behavior of the compared methods. PG-ML gives the highest structure-background contrast-to-noise ratio (CNR) because of its low background standard deviation, but it also has the lowest structural contrast among the baselines and the proposed method. In contrast, the proposed method gives the highest structural contrast while maintaining a low background fluctuation and a high CNR. This indicates that the improvement is not limited to background suppression, but also involves better retention of structural intensity variations. The PSD analysis in Fig. 3i supports this interpretation. The proposed method preserves the largest mid- and high-frequency PSD energies, whereas PG-ML shows reduced high-frequency content, consistent with partial smoothing of fine structural features.

## 2.4 Computational efficiency and reconstruction stability

Beyond imaging performance and noise suppression, practical deployment of the proposed framework also depends critically on computational efficiency and stability. Iterative phase retrieval in ptychography is typically computationally demanding because of the large data volumes and repeated Fourier transforms involved. To assess this aspect, we compared the computational cost of the proposed framework with conventional strategies, as shown in Figs. 4a,c. By embedding noise suppression into the in-the-loop projection step, pixelwise weighting prioritizes diffraction-rich regions and guides iterative updates toward informative structures, thereby accelerating convergence. As a result, the algorithm recovers the correct phase values more easily and converges more rapidly under the same number of iterations. In practice, convergence is achieved within approximately 200 iterations regardless of imaging mode, and reconstructions obtained after 300 iterations already exhibit high clarity that is nearly indistinguishable from those obtained after 500 iterations.

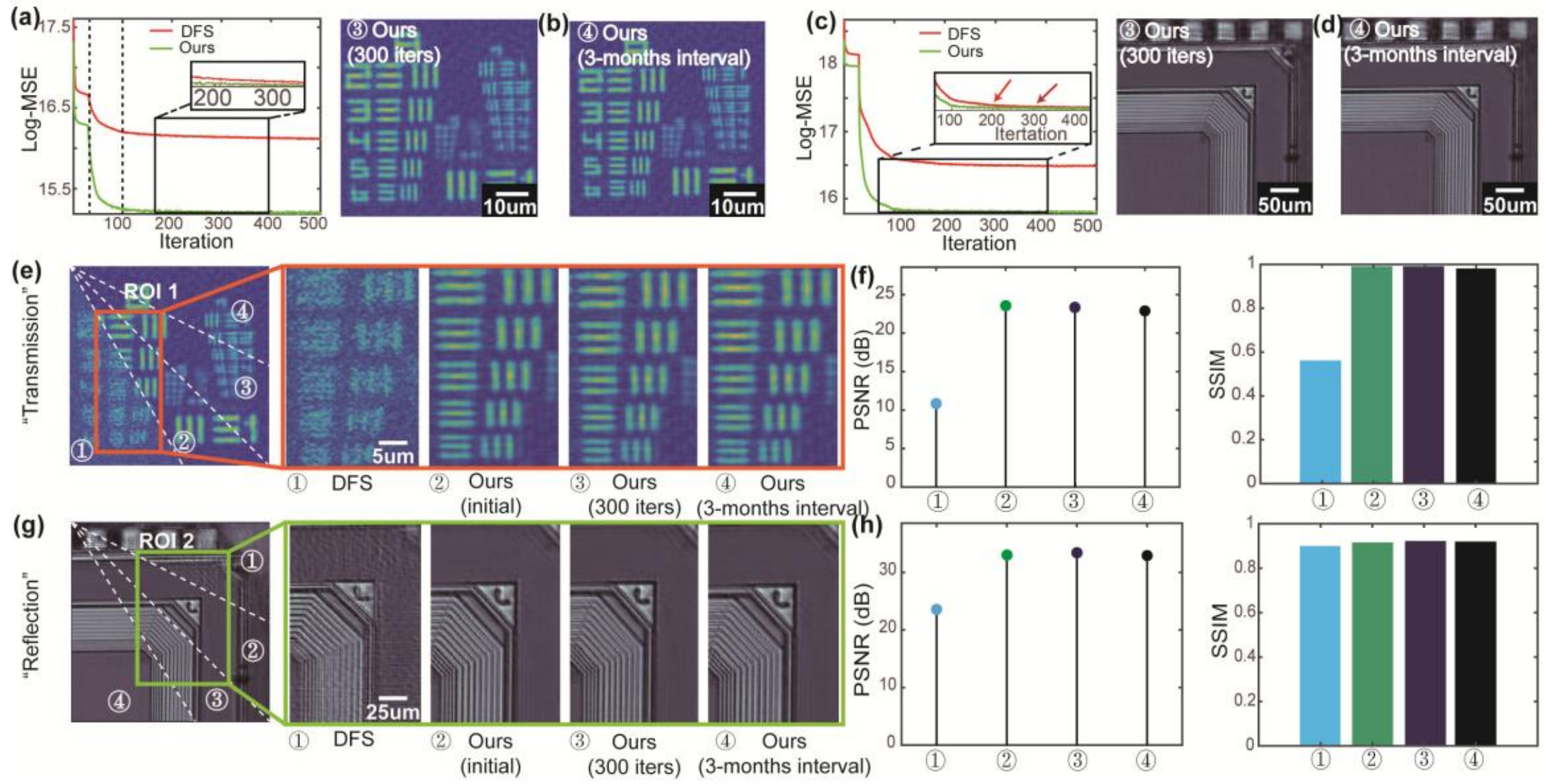

*Fig. 4 Evaluation of computational efficiency and reconstruction stability of the proposed framework. (a), (c) Comparison of convergence behavior and loss evolution between the proposed method and conventional denoising strategies under two imaging modes. (b), (d) Reconstruction results obtained after a three-month interval using the same pre-calibrated denoising framework. (e)-(f) Statistical comparison across four imaging strategies for the transmission-mode test sample (ROI 1), where reconstruction quality is quantitatively evaluated using PSNR and SSIM metrics. (g)-(h) Quantitative comparison for the reflection-mode test sample (ROI 2) based on PSNR and SSIM analysis.*

To further assess scalability and stability, we repeated the acquisition and reconstruction of two samples after an interval of three months while reusing the previously calibrated denoising framework. The results, shown in Figs. 4b,d, consistently preserved high resolution and maintained uniform intensity variations across both background and structural regions. This long-term stability arises from the design of the framework, which abandons conventional prior-based models and instead relies on confidence measures derived directly from diffraction features. Since these feature-driven weights do not vary with time, the framework retains its effectiveness across repeated acquisitions, confirming its robustness for practical deployment. To eliminate the subjectivity of qualitative inspection, we employed peak signal-to-noise ratio (PSNR) and structure similarity index measure (SSIM) as quantitative metrics of reconstruction quality. As summarized in Fig. 4e, for the transmission-mode test sample, ROI 1 covering all elements in Group 8 was selected. The quantitative evaluation across four imaging strategies is reported in Fig. 4f. The DFS achieves a PSNR of only about 11 dB, whereas the other three modes exhibit at least a two-fold improvement. Moreover, both at 300 iterations and under reuse of the denoising framework after three months, the SSIM values consistently approach unity, directly confirming the efficiency and stability of the proposed approach in practical deployment. Additional experiments in reflection geometry, presented in Figs. 4g,h, confirm that this property is preserved. Although the structural similarity of DFS reaches about 0.9, the reconstructed images still suffer from noise interference and blurred etched regions, leading to PSNR values far below those of the other three methods.

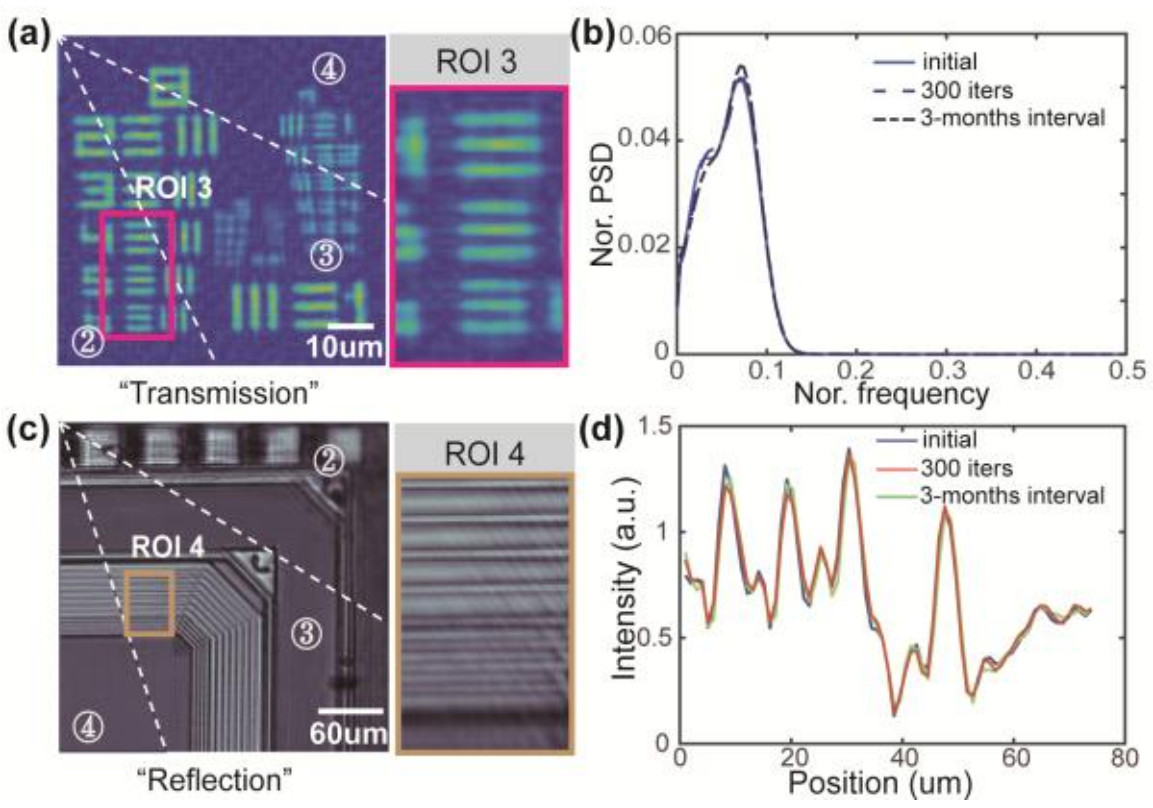

*Fig. 5 Quantitative evaluation of spectral and intensity stability across reconstruction conditions. (a)–(b) PSD comparison among the initial reconstruction, the 300-iteration result, and the re-acquired data after three months, showing identical peak positions at 0.07 pixel$^{-1}$ and consistent high-frequency decay. (c)–(d) Intensity-distribution analysis of the reflection-mode sample under the same three reconstruction states, confirming consistent spatial contrast and stability across time.*

The aforementioned quantitative evaluations, while informative, cannot fully reflect the stability of the proposed framework, since they rely only on region-specific metrics within individual reconstructions. To provide a more comprehensive assessment, we further compared the power spectral density (PSD) and intensity statistics across identical regions obtained under three conditions: the initial state, after 300 iterations, and following a three-month interval with reused calibration. As shown in Figs. 5a,b, the PSD curves of the three methods remain nearly identical, with the main peak located at a spatial frequency of 0.07 pixel$^{-1}$ and exhibiting the same amplitude, while the high-frequency components ($> 0.1$) rapidly decay to zero in all cases. Likewise, for ROI 4 in the reflection-mode semiconductor sample, the intensity profiles presented in Fig. 5d exhibit excellent overlap across the three settings. Taken together, both the consistently high quantitative indicators and the cross-condition agreement confirm that the proposed framework achieves practical efficiency and stability.

## 2.5 Generalization to complex amorphous biological structures

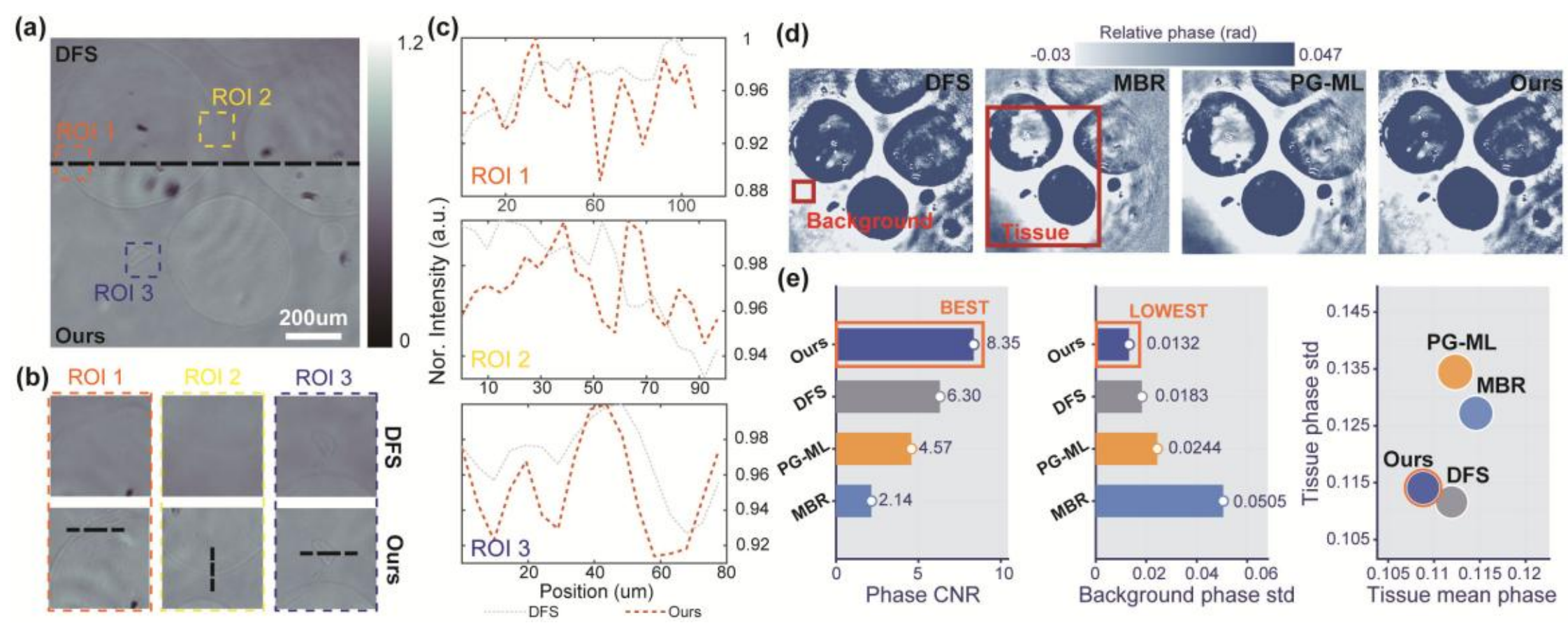

*Fig. 6 Experimental validation on a hybrid carp ovary tissue section. (a) Reconstruction comparison of a local tissue region obtained with DFS and the proposed method. (b) Magnified views of three selected ROIs. ROI 1 and ROI 3 contain tissue structures, while ROI 2 corresponds to a relatively homogeneous region. (c) Cross-sectional intensity profiles extracted from the ROIs in (b). (d) Relative phase reconstructions obtained with DFS, MBR, PG-ML, and the proposed method. The red boxes indicate the tissue and background regions used for phase-statistics analysis. (e) Quantitative comparison of tissue-background phase CNR, background phase standard deviation, and tissue phase statistics.*

The USAF target and semiconductor sample provide well-defined edges for evaluating resolution and anisotropic reconstruction behavior. To assess performance on weakly scattering objects with irregular morphology, we further imaged a hybrid carp ovary tissue section. Such biological samples contain continuous phase variations rather than binary line structures, making the reconstruction more sensitive to residual background fluctuations and low-contrast phase features. Fig 6a compares a representative tissue region reconstructed by DFS and by the proposed method. The DFS result retains the overall tissue outline, but fine variations within the cell-like structures are partially masked by background texture. The proposed reconstruction gives a more stable background and clearer tissue boundaries, especially in regions where the contrast is weak.

Three ROIs were selected for local analysis, as shown in Fig. 6b. ROI 1 and ROI 3 contain weak tissue features, whereas ROI 2 was chosen from a relatively uniform region to evaluate background stability. The intensity profiles in Fig. 6c show that the proposed method follows the local structural variations more clearly than DFS. In ROI 3, for example, the proposed profile preserves the valley around 40–60 μm, while the DFS profile is flatter and more affected by low-amplitude fluctuations. In ROI 2, the proposed profile remains more uniform, consistent with reduced background variation.

We further compared the relative phase reconstructions obtained with DFS, MBR, and PG-ML, as shown in Fig. 6d. Owing to the weak-scattering nature of the biological tissue, the diffraction information is mainly distributed as faint ring-like features around the illumination footprint rather than as isolated high-contrast diffraction orders. MBR and PG-ML do not explicitly distinguish these weak tissue-related diffraction features from detector background, which leads to visible fluctuations in the reconstructed tissue region. In the selected tissue , part of the object boundary is inevitably included, resulting in a higher tissue mean for MBR and PG-ML. We therefore used the tissue-background phase CNR to jointly reflect tissue-background separation and residual background fluctuation. As shown in Fig. 6e, the proposed method gives the highest phase CNR and the lowest background phase standard deviation, indicating improved visibility of weak tissue features while suppressing background-induced phase variations.

# 3 Discussion

We have developed a detector-informed confidence-weighted reconstruction framework for ptychographic imaging with non-ideal detection hardware. Rather than treating the detector as a spatially uniform measurement plane, the method incorporates calibrated pixelwise reliability into the amplitude constraint. The confidence map $w(\boldsymbol{q}, t_{\mathrm{exp}})$ controls the contribution of detector-plane residuals to the exit-wave correction, so that unreliable detector responses are down-weighted within the reconstruction loop instead of being removed as a separate preprocessing step.

The comparison with DFS, MBR, and PG-ML clarifies this distinction. DFS and MBR reduce background before reconstruction, but the corrected diffraction data are then used with uniform confidence. PG-ML introduces a prescribed Poisson–Gaussian noise model into the reconstruction loop, which can strongly suppress background fluctuations but may also attenuate weak structural variations when the actual detector response departs from the assumed model. In contrast, the proposed method uses calibrated detector reliability to regulate the measurement residual itself, allowing background-induced contributions to be reduced while retaining physically meaningful diffraction information.

Across the three experimental settings, this behavior led to consistent improvements in reconstruction fidelity. In the transmission USAF experiment, the proposed method preserved line-pair modulation and reached a $k$-factor of approximately 0.65, close to the Rayleigh criterion. In reflection geometry, it reduced anisotropic artifacts caused by the tilted diffraction geometry and preserved stronger mid- and high-frequency structural components than the compared baselines. For the weakly scattering biological tissue, it achieved the highest tissue-background phase CNR and the lowest background phase fluctuation, indicating improved visibility of weak phase features.

The present implementation remains limited by the range and quality of the detector calibration. In the reflection experiment, the exposure time exceeded the calibrated range, so the confidence map relied partly on extrapolated detector-response information. In addition, photon-free calibration mainly characterizes detector-originated background fluctuations and does not fully account for signal-dependent shot noise. Future work could combine detector-response calibration with intensity-dependent uncertainty estimation to further improve robustness under broader exposure and illumination conditions. Overall, these results show that detector reliability can serve as an effective in-loop measurement constraint for ptychographic reconstruction. The method does not replace the measured diffraction data or impose an external image prior; instead, it weights the measurement residual according to calibrated detector confidence. This principle may also be useful for other coherent imaging modalities governed by intensity measurements, provided that detector-response information can be incorporated into the corresponding data-fidelity term.

# 4 Methods

## 4.1 Experimental configuration and data acquisition

To comprehensively validate the universality of the proposed framework, we conducted experiments across three representative sample types, covering both transmission and reflection geometries. In transmission mode, two distinct samples were imaged to test amplitude and phase recovery capabilities: a commercial high-resolution USAF-1951 test target (Newport) with a minimum feature size of 137 nm, serving as a standard binary amplitude object; and an unstained tissue section of a hybrid carp ovary, representing complex biological structures characterized by continuous, low-contrast phase variations. In reflection mode, a custom-fabricated microstructured semiconductor chip was employed to evaluate the framework's robustness against anisotropic diffraction artifacts induced by oblique illumination. Reference images were acquired using a standard optical microscope to facilitate morphological verification. All sample preparation and measurements were performed in a strictly controlled cleanroom environment.

The experimental setup was based on a home-built optical microscope system (Fig. 1a). To manage the high dynamic range of the diffraction signal and prevent saturation of the zero-order beam, neutral-density filters (ND1 and ND2) were employed in conjunction with optimized camera exposure times. Samples were mounted on a custom-designed holder attached to a high-precision linear translation stage. To maximize the numerical aperture (NA) and captured spatial frequency information, the detector was positioned at the minimum mechanically permissible distance from the sample.

Data acquisition was executed via a LabVIEW-based automated control interface. The translation stage followed a Fermat spiral trajectory (Fig. 1b) to ensure uniform coverage and overlap. The scan comprised 321 positions for transmission experiments and 302 for reflection experiments, with a nominal average step size of 10 μm. In reflection geometry, the oblique incidence angle resulted in an elliptical illumination footprint; while the nominal step size was maintained, the effective overlap ratio was naturally reduced. A stabilization delay of 500 ms was implemented after each stage movement prior to detector exposure to minimize mechanical vibration artifacts.

## 4.2 Detector-confidence-weighted reconstruction framework

The reconstruction pipeline was built upon the momentum-accelerated ptychographic iterative engine (mPIE) (Andrew Maiden et al. 2017), implemented in MATLAB. As detailed in Supplementary Note 1, the measured diffraction intensity at the $j$-th probe position can be written in compact form as

$$I_j^{\mathrm{mea}}(\boldsymbol{q}) = |A_j x|^2 + n_j(\boldsymbol{q}),$$

where $I_j^{\mathrm{mea}}(\boldsymbol{q})$ is the measured intensity at spatial frequency coordinate $\boldsymbol{q}$, $A_j$ represents the forward operator including probe modulation, lateral translation, and Fourier propagation, $x$ denotes the vectorized object function, and $n_j(\boldsymbol{q})$ collects perturbations introduced along the acquisition chain, including illumination fluctuations, speckle noise, and detector-related readout errors. Standard reconstruction algorithms usually treat the detector as an ideal measurement plane and therefore apply the measured intensity as an equally reliable constraint at all spatial frequency coordinates.

To account for the nonuniform and exposure-dependent detector response, we construct a spatially resolved confidence map $w(\boldsymbol{q}, t_{\mathrm{exp}})$ from calibrated sensor-response characteristics. Here, $t_{\mathrm{exp}}$ denotes the camera exposure time. It is derived from the photon-free camera-noise calibration procedure reported in our previous work (Xie, Zhou, et al. 2025), in which the pixelwise offset and variance are modeled as functions of exposure time. After cross-exposure modeling and background-offset removal, Gaussian smoothing is used to capture the spatial nonuniformity of photon and noise responses while reducing isolated pixel-level fluctuations, as described in Supplementary Note 3. Controlled ablation experiments removing extended cross-exposure modeling or Gaussian smoothing are provided in Supplementary Fig. 4, where both variants show lower structure-background CNR and stronger background fluctuation than the full method.

In the conventional amplitude-projection step, the measured amplitude is imposed uniformly,

$$\Psi_j^{\mathrm{std}}(\boldsymbol{q}) = \sqrt{I_j^{\mathrm{mea}}(\boldsymbol{q})}\,\frac{\Psi_j(\boldsymbol{q})}{|\Psi_j(\boldsymbol{q})| + \varepsilon},$$

where $\Psi_j(\boldsymbol{q})$ is the current detector-plane wavefield estimate and $\varepsilon$ is a small positive constant used to avoid division by zero. This update assumes that the amplitude residual at each spatial frequency coordinate is equally trustworthy. In the proposed formulation, the measured amplitude is instead imposed with a pixel-dependent reliability weight,

$$\Psi_j'(\boldsymbol{q}) = \left[(1 - w(\boldsymbol{q}, t_{\mathrm{exp}}))|\Psi_j(\boldsymbol{q})| + w(\boldsymbol{q}, t_{\mathrm{exp}})\sqrt{I_j^{\mathrm{mea}}(\boldsymbol{q})}\right]\frac{\Psi_j(\boldsymbol{q})}{|\Psi_j(\boldsymbol{q})| + \varepsilon},$$

where $\Psi_j'(\boldsymbol{q})$ is the detector-plane wavefield after the confidence-weighted amplitude update. When $w(\boldsymbol{q}, t_{\mathrm{exp}}) = 1$, the update reduces to the conventional amplitude projection. When $w(\boldsymbol{q}, t_{\mathrm{exp}}) = 0$, the current predicted amplitude is retained at that spatial frequency coordinate. For $0 < w(\boldsymbol{q}, t_{\mathrm{exp}}) < 1$, the measured amplitude is enforced in proportion to the calibrated detector reliability.

This update can be written equivalently in terms of the spatial frequency amplitude residual. Let

$$r_j(\boldsymbol{q}) = \sqrt{I_j^{\mathrm{mea}}(\boldsymbol{q})} - |\Psi_j(\boldsymbol{q})|$$

denote the difference between the measured and predicted amplitudes. The confidence-weighted projection is then

$$\Psi_j'(\boldsymbol{q}) = \Psi_j(\boldsymbol{q}) + w(\boldsymbol{q}, t_{\mathrm{exp}})r_j(\boldsymbol{q})\frac{\Psi_j(\boldsymbol{q})}{|\Psi_j(\boldsymbol{q})| + \varepsilon}.$$

Thus, $w(\boldsymbol{q}, t_{\mathrm{exp}})$ controls the fraction of the detector-plane residual that is allowed to enter the iterative correction. The same weighting can be expressed through the amplitude-consistency term

$$\mathcal{L}_{\mathrm{amp}} = \frac{1}{2}\sum_{j,\boldsymbol{q}} w\,(\boldsymbol{q}, t_{\mathrm{exp}})\left(|\Psi_j(\boldsymbol{q})| - \sqrt{I_j^{\mathrm{mea}}(\boldsymbol{q})}\right)^2.$$

This formulation is mathematically related to weighted least-squares reconstruction, in which measurements with different reliability levels are assigned different residual weights. In a conventional weighted least-squares model, the weight is often chosen according to an assumed or calibrated measurement variance, for example $w_i \propto 1/\sigma_i^2$. In the present implementation, $w(\boldsymbol{q}, t_{\mathrm{exp}})$ plays an analogous role as a pixelwise reliability coefficient for the detector-plane amplitude constraint. The distinction is that $w(\boldsymbol{q}, t_{\mathrm{exp}})$ is not prescribed from a global analytical noise law. Instead, it is constructed from calibrated, spatially and exposure-resolved sensor-response characteristics and then applied within the ptychographic amplitude-projection step. Thus, the proposed method can be viewed as a detector-informed weighted residual formulation, where the measurement confidence is derived from the physical response of the detector and used to regulate the iterative update.

The effect of this weighting is carried into the object and probe updates through the corrected exit wave. Let

$$\Delta\psi_j^w(\boldsymbol{r}) = \psi_j'(\boldsymbol{r}) - \psi_j(\boldsymbol{r})$$

denote the confidence-weighted exit-wave correction, where $\psi_j(\boldsymbol{r})$ and $\psi_j'(\boldsymbol{r})$ are the exit waves before and after the amplitude update, respectively, and $\boldsymbol{r}$ is the real-space coordinate. The corresponding mPIE updates are written as

$$O'(\boldsymbol{r}) = O(\boldsymbol{r}) + \alpha \frac{P^*(\boldsymbol{r}-\boldsymbol{r}_j)}{\max(|P(\boldsymbol{r}-\boldsymbol{r}_j)|^2)} \Delta\psi_j^w(\boldsymbol{r}),$$
$$P'(\boldsymbol{r}-\boldsymbol{r}_j) = P(\boldsymbol{r}-\boldsymbol{r}_j) + \beta \frac{O^*(\boldsymbol{r})}{\max(|O(\boldsymbol{r})|^2)} \Delta\psi_j^w(\boldsymbol{r}),$$

where $O(\boldsymbol{r})$ is the object function, $P(\boldsymbol{r}-\boldsymbol{r}_j)$ is the probe shifted to the scan position $\boldsymbol{r}_j$, $\alpha$ and $\beta$ are the object and probe update parameters, and $(\cdot)^*$ denotes complex conjugation.

These equations show how the detector confidence map enters the reconstruction. The object and probe are not weighted directly; instead, their increments are driven by the exit-wave correction generated from the weighted detector-plane residual. For interpretation, the residual may be regarded as containing a diffraction-related part and a detector-induced part,

$$r_j(\boldsymbol{q}) = r_j^{\mathrm{diff}}(\boldsymbol{q}) + r_j^{\mathrm{det}}(\boldsymbol{q}),$$

where $r_j^{\mathrm{diff}}(\boldsymbol{q})$ denotes the component associated with physically meaningful diffraction information and $r_j^{\mathrm{det}}(\boldsymbol{q})$ denotes the component associated with unreliable detector response. The weighted residual becomes

$$w(\boldsymbol{q}, t_{\mathrm{exp}})r_j(\boldsymbol{q}) = w(\boldsymbol{q}, t_{\mathrm{exp}})r_j^{\mathrm{diff}}(\boldsymbol{q}) + w(\boldsymbol{q}, t_{\mathrm{exp}})r_j^{\mathrm{det}}(\boldsymbol{q}).$$

In detector regions assigned high confidence, the residual contribution is largely retained. In regions assigned low confidence, the contribution of detector-induced residuals is reduced before it contributes to $\Delta\psi_j^w(\boldsymbol{r})$ and, consequently, to the updates of $O$ and $P$. The proposed formulation therefore acts within the iterative amplitude constraint rather than as a post-reconstruction correction or as a replacement of the measured diffraction data. It provides a detector-confidence-weighted extension of the conventional ptychographic update, in which measured diffraction information is used according to its calibrated spatial and exposure-dependent reliability.

## Author Contributions

“Yun Xie conceived the study, developed the methodology, performed the experiments and data analysis.”

“Bianli Zhao and Rui Zhang contributed to experimental maintenance and system stability.”

“Han Yue assisted with language checking and manuscript polishing.”

“Zhiyi Huang and Weiran Jiang assisted with data processing.”

“Chuangchuang Cheng contributed to sample preparation and experimental discussion.”

“Steve F. Shu supervised the project, provided guidance, secured funding, and reviewed the manuscript.”

## Conflicts of Interest

The authors declare no competing financial or non-financial interests.

## Data Availability

The datasets generated and analyzed during the current study are not publicly available due to project restrictions but are available from the corresponding author upon reasonable request.

## Code Availability

Codes used to post-process the diffraction data with in this paper are available from the reasonable request.

# Supplementary Materials

Supplementary Figs. 1 to 4